\documentclass[a4paper]{jpconf}

\usepackage{amsmath,amsthm,amsfonts}
\usepackage[utf8x]{inputenc}
\begin{document}

\title{Inhomogenous loop quantum cosmology with matter}
\author{D Martín-de~Blas${}^{1}$, M Martín-Benito${}^{2}$ and G A Mena Marugán${}^{1}$}
\address{${}^{1}$ Instituto de Estructura de la Materia, IEM-CSIC, Serrano 121, 28006 Madrid, Spain.}
\address{${}^{2}$ MPI f\"ur Gravitational Physics, Albert Einstein Institute,
Am M\"uhlenberg 1, D-14476 Potsdam, Germany.}
\ead{daniel.martin@iem.cfmac.csic.es, mercedes@aei.mpg.de, mena@iem.cfmac.csic.es}

\begin{abstract}
The linearly polarized Gowdy $T^3$ model with a massless scalar
field with the same symmetries as the metric is quantized by applying a hybrid approach. The
homogeneous geometry degrees of freedom are loop quantized, fact which leads to
the resolution of the cosmological singularity, while a Fock quantization is employed for both
matter and gravitational inhomogeneities. Owing to the inclusion of the massless scalar
field this system allows us to modelize flat Friedmann-Robertson-Walker cosmologies filled with
inhomogeneities propagating in one direction. It provides a perfect scenario to study
the quantum back-reaction between the inhomogeneities and the polymeric homogeneous and isotropic
background.
\end{abstract}

\section{Introduction}
Loop quantum cosmology (LQC) \cite{lqc1,lqc2,lqc3} is a quantization for cosmological systems
based on ideas and techniques of loop quantum gravity (LQG) \cite{lqg}. Several homogeneous
cosmologies have been satisfactory quantized within LQC \cite{lqc2}. The loop
quantization leads to a resolution of the cosmological singularities, that
turn out to be replaced by quantum bounces \cite{aps}. 

In order to study inhomogeneous cosmological systems in the framework of LQC, a hybrid quantization
has been developed and applied to the Gowdy $T^3$ model with linear polarization
\cite{hybrid1,hybrid2,hybrid3,hybrid4}. This hybrid quantization applies to reduced models in which
only global constraints remain to be imposed at the quantum level. It combines the
polymeric quantization of LQC for the homogeneous degrees of freedom with a Fock quantization
for the inhomogeneities, so that one can deal with the field complexity.
Therefore, this hybrid procedure assumes that there exists a hierarchy in the loop quantum geometry
effects such that the most relevant ones are those associated with the homogeneous degrees of
freedom. 

With the aim of studying more physically interesting models we introduce a massless scalar field in
the linearly polarized Gowdy $T^{3}$ model \cite{hybrid5}. This field is minimally coupled and has
the same symmetries of the geometry. Like in the vacuum model, the application of the hybrid
quantization turns out in the resolution of the initial singularity, achieved thanks to the
particular action of the Hamiltonian constraint whose homogeneous terms are loop quantized, and in
the recovering of the standard Fock quantization for the inhomogeneities. 
The interest of the current model lies in the fact that, unlike the
vacuum case, it admits as subset of homogeneous an isotropic solutions the flat
Friedmann-Robertson-Walker (FRW) model, so that we can regard the system as an approximated flat-FRW background
filled with inhomogeneities propagating in one direction. Therefore this model is perfect to
analyze the quantum effects of anisotropies and inhomogeneities on a flat FRW background that is loop quantized, and vice versa, to study how the polymeric
background affects the evolution of the anisotropies and inhomogeneities. Actually, we are now
developing approximations to obtain physically interesting solutions of the Hamiltonian constraint \cite{lrsg}. 

The Gowdy $T^{3}$ model with linear polarization presents a subset of classical solutions with local
rotational symmetry (LRS). To simplify the calculations one can work with the reduced
\emph{LRS-Gowdy model} without loss of generality. We will see how to implement this reduction at
the quantum
level using the map introduced in Ref. \cite{awe}. 
 
\section{Hybrid Quantization}
We apply a hybrid quantization to the linearly polarized Gowdy $T^{3}$ model with a minimally
coupled massless scalar field. These cosmologies are globally hyperbolic spacetimes with spatial
three torus topology and with two axial and hypersurface orthogonal Killing vectors fields
$\partial_\sigma$ and $\partial_\delta$. We choose coordinates $\{t,\theta,\sigma,\delta\}$ adapted
to the symmetries such that all our fields only depend on time and on the spatial coordinate
$\theta \in S^1$. After performing a symmetry reduction and a partial gauge fixing as
in Ref. \cite{hybrid2}, we obtain a reduced phase space that can be split in two sectors by
expanding the fields in Fourier modes. The zero modes form the \emph{homogeneous sector}, which is
equivalent to the phase space of the Bianchi I model with an homogeneous massless scalar field,
$\phi$. The \emph{inhomogeneous sector} contains the non-zero modes of both gravitational and matter
fields, $\xi$ and $\varphi$ respectively. With the aim of quantizing the homogeneous sector of the
geometry by using LQC techniques, we describe it in terms of the Ashtekar-Barbero variables of the
Bianchi I model with three-torus topology. In a diagonal gauge these variables are given by the
three components of the
densitized triad $p_{j}$ and of the $su(2)$-connection $c_{j}$, with $j=\theta, \sigma,
\delta$. They satisfy $\{c_{i}, p_{j}\}=8\pi \gamma G \delta_{ij}$, where $\gamma$ is the
Immirzi parameter and $G$ is the Newton constant.
It has been proven that the
deparametrized model admits a unique satisfactory Fock quantization \cite{cmmv}. Then we describe
the inhomogeneities employing the annihilation and creation variables chosen in \cite{cmmv}.
Two global constraints remain to be imposed at the quantum level, a momentum
constraint $\mathcal C_{\theta}$ that generates translations in the circle and only involves the
inhomogeneous sector, and the Hamiltonian constraint,
$\mathcal{C}=\mathcal{C}_{\text{hom}}+\mathcal{C}_{\text{inh}}$, formed by a homogeneous term that
is not other thing but the Hamiltonian constraint of the Bianchi I model with a homogeneous scalar
$\phi$, and by an additional term that couples homogeneous and inhomogeneous sectors.

In order to quantize the system we have to define a representation of the basic variables on a kinematical Hilbert space, that in our case takes this form:
 $\mathcal{H}_{\text{kin}}=\mathcal{H}_{\text{kin}}^{\text{BI}}\otimes L^{2}(\mathbb{R},d\phi)\otimes \mathcal{F}^{\xi}\otimes \mathcal{F}^{\varphi}$.
Here $\mathcal{H}_{\text{kin}}^{\text{BI}}$ is the kinematical Hilbert space for the Bianchi
I model within the improved dynamics of LQC \cite{awe},
$\mathcal{F}^{\xi}$ and $\mathcal{F}^{\varphi}$ are identical Fock spaces, for the gravitational
waves and for the matter inhomogeneities respectively, and $L^{2}(\mathbb{R},d\phi)$ is
the Hilbert space accounting for the homogeneous scalar $\phi$ for which we employ a
standard Schr\"odinger representation. The construction of the kinematical Hilbert space
$\mathcal{H}_{\text{kin}}^{\text{BI}}$ in LQC mimics that of LQG in the sense that the connection
is not defined in the quantum theory but only its holonomies. The inner
product is discrete, so that, the operators $\hat{p}_{j}$ have a
discrete spectrum equal to the real line, and their mutual eigenstates
$|p_\theta,p_\sigma,p_\delta\rangle$ form a orthonormal basis of
$\mathcal{H}_{\text{kin}}^{\text{BI}}$. Because of the improved dynamics, the action of the basic
holonomy operators, $\hat{\mathcal{N}}_{\pm\bar{\mu}_{j}}$, on the states
$|p_\theta,p_\sigma,p_\delta\rangle$ is quite involved. It is convenient to relabel them as
$|\lambda_{\theta},\lambda_{\sigma},v\rangle$, where $\lambda_{j}\propto
\text{sgn}(p_{j})\sqrt{|p_{j}|}$ and $v=2\lambda_{\theta}\lambda_{\sigma}\lambda_{\delta}$
is proportional to the Bianchi I volume, or \emph{homogeneous} volume.
The holonomy operators $\hat{\mathcal{N}}_{\pm\bar{\mu}_{j}}$ scale the label
$\lambda_{j}$ in such a way that the label $v$ is simply shifted by one \cite{awe}. For the Fock
spaces, $\mathcal{F}^{\alpha} 
(\alpha=\xi,\varphi)$, an orthonormal basis is given by the \emph{n}-particle states. We call
$n_m^{\alpha}$ the occupation number of the field $\alpha$ in the mode $m$. The
creation-annihilation variables are promoted to creation-annihilation operators, denoted by
$\hat{a}_{m}^{(\alpha)\dagger}, \hat{a}_m^{(\alpha)}$, respectively.

Now we can construct the operators that represent the
constraints. The generator of the translations in the circle is represented by the operator
\begin{equation}
 \widehat{\mathcal{C}}_{\theta}=\sum_{m=1}^{\infty}m\left(\widehat{X}^{\xi}_{m}+\widehat{X}^{\varphi
}_{m}\right), 
\qquad\widehat{X}_{m}^{\alpha}=\hat{a}^{(\alpha)\dagger}_{m}\hat{a}^{(\alpha)}_{m}-\hat{a}^{
(\alpha)\dagger}_{-m}\hat{a}^{(\alpha)}_{-m}.
\end{equation}
The imposition of this constraint on
$\mathcal{F}^{\xi}\otimes\mathcal{F}^{\varphi}$ leads to the condition,
$\sum_{m=1}^{\infty}m(X^{\xi}_{m}+ X^{\varphi}_{m})=0$, where
$X^{\alpha}_{m}=n^{\alpha}_{m}-n^{\alpha}_{-m}$. The $n$-particle states that satisfy this
condition form a proper Fock subspace $\mathcal{F_{\text{p}}}\subset \mathcal{F}^{\xi}\otimes
\mathcal{F}^{\varphi}$ that is unitarily equivalent to the physical space of the Fock quantization
of the deparametrized model, obtained as in vacuum case \cite{cmmv}.

For the Hamiltonian constraint operator we follow the construction of the vacuum case
\cite{hybrid3}, so that the operator decouples the states of zero homogeneous volume and does not
relate states with different sign of any of the homogeneous geometry labels
$\lambda_{\theta}, \lambda_{\sigma}$ and $v$. Consequently, we can remove from our theory the states
which are the analog of the classical singularity (those with $v=0$) and restrict the study
to the sector with strictly positive labels $\lambda_{\theta}, \lambda_{\sigma}$ and $v$, for
instance. The Hamiltonian constraint operator reads,
\begin{equation}
\widehat{\mathcal{C}}=-\sum_{i\neq j}\sum_{j}\dfrac{\widehat{\Theta}_{i}\widehat{\Theta}_{j}} {16\pi G \gamma^{2}}-\dfrac{\hbar^{2}}{2}\left[\dfrac{\partial}{\partial \phi}\right]^{2}\!\!+ 2\pi\hbar\widehat{|p_{\theta}|}\widehat{H}_{0}+\hbar\widehat{\left[\dfrac{1}{|p_{\theta}|^{\frac{1}{4}}}\right]}^{2}\!\dfrac{\left(\widehat{\Theta}_{\delta}+\widehat{\Theta}_{\sigma}\right)^{2}}{16\pi \gamma^{2}}\widehat{\left[\dfrac{1}{|p_{\theta}|^{\frac{1}{4}}}\right]}^{2}\!\!\widehat{H}_{\text{int}},
\end{equation}
with $i,j \in \{\theta,\sigma,\delta\}$. Here we have introduced the quantum version of
${c_{j}p_{j}}$, given by
\begin{equation}
\widehat{\Theta}_{j}=i\pi \gamma  G \hbar
\widehat{\sqrt{|v|}}\hspace*{-0.5mm}\left[\hspace*{-0.5mm}\left(\hspace*{-0.5mm}\hat{\mathcal{N}}_{
-2\bar{\mu}_{j}}\hspace*{-0.5mm}-\hat{\mathcal{N}}_{2\bar{\mu}_{j}}\hspace*{-0.5mm}\right)\widehat{
\text{sgn}(p_{j})}\hspace*{-0.5mm}+\widehat{{\text{sgn}}(p_{j})}\left(\hspace*{-0.5mm}\hat{\mathcal{
N}}_{-2\bar{\mu}_{j}}\hspace*{-0.5mm}-\hat{\mathcal{N}}_{2\bar{\mu}_{j}}\hspace*{-0.5mm}
\right)\right]\widehat{\sqrt{|v|}},
\end{equation}
and the operator $\widehat{\left[1/|p_{\theta}|^{1/4}\right]}$, which is regularized
as usual by taking the commutator of some suitable power of $p_\theta$ with the holonomies
\cite{hybrid3}. The operators $\widehat{H}_{\text{int}}$ and $\widehat{H}_{0}$ act non-trivially
only on the inhomogeneous sector of the kinematical Hilbert space and are given by
\begin{equation}
\widehat{H}_{0}=\!\!\!\sum_{\alpha \in \{\xi,\varphi\}}\sum_{m=1}^{\infty}m
\widehat{N}^{\alpha}_{m}, \qquad
\widehat{H}_{\text{int}}=\!\!\!\sum_{\alpha \in \{\xi,\varphi\}}
\sum_{m=1}^{\infty}\dfrac{1}{m}\left(\widehat{N}^{\alpha}_{m}+
\hat{a}^{(\alpha)\dagger}_{m}
\hat{a}^{(\alpha)\dagger}_{-m} + \hat{a}^{(\alpha)}_{m}
\hat{a}^{(\alpha)}_{-m}\right),
\end{equation}
where $\widehat{N}^{\alpha}_{m}=\hat{a}^{(\alpha)\dagger}_{m}
\hat{a}^{(\alpha)}_{m} + \hat{a}^{(\alpha)\dagger}_{-m}
\hat{a}^{(\alpha)}_{-m}$. Note that the inhomogeneities of both fields contribute to the constraints in
exactly the same way.

The Hamiltonian constraint operator does not relate all states with different values of $v$ and
$\lambda_j$ $(j= \theta,\sigma)$, but \emph{superselects} different sectors. The superselection
sectors in $v$ are semilattices of step four in $\mathbb{R}^{+}$ from a minimum value $\epsilon \in
(0,4]$. The superselection sectors in $\lambda_j$ $(j= \theta,   \sigma)$ are more
involved. Given initial data $\lambda_{j}^{\ast}$ and $\epsilon$, the values of $\lambda_j$ in the
corresponding sector are of the form $\lambda_j=\omega_{\epsilon}\lambda_{j}^{\ast}$, where
$\omega_{\epsilon}$ runs over a certain countable set, that is dense in $\mathbb{R}^{+}$ (see
\cite{hybrid3}).

The quantum Hamiltonian constraint leads to a difference equation in the variable $v$, and then we
can regard it as an evolution equation in $v$. As in the vacuum model \cite{hybrid3},
the solutions are completely determined by a dense set of data in the
initial section $v=\epsilon$. This property allows us to characterize the
physical Hilbert space as the Hilbert space of these initial data, whose inner product is
determined imposing reality conditions in a complete set of observables. The result is
$\mathcal{H}_{\text{phys}}=\mathcal{H}_{\text{phys}}^{\text{BI}}\otimes
L^{2}(\mathbb{R},d\phi) \otimes \mathcal{F}_{\text{p}}$, where $\mathcal{H}_{\text{phys}}^{\text{BI}}$ is the physical Hilbert space for Bianchi I cosmologies given in \cite{hybrid4}.

\section{{\emph{Projection}} to LRS-Gowdy Model}
The linearly polarized Gowdy $T^{3}$ model is
symmetric under the interchange $\sigma \longleftrightarrow \delta$. Therefore, there is a subset of
classical solutions with local rotational symmetry and we will call LRS-Gowdy model the system that
they define. The quantum LRS-Gowdy model can be obtained from the quantum Gowdy model that we have
constructed above by using the map 
\begin{equation}
|\Psi(\lambda_{\theta},\lambda_{\sigma}, v)\rangle \qquad \longrightarrow
\qquad \sum_{\lambda_{\sigma}}|\Psi(\lambda_{\theta},\lambda_{\sigma}, v)\rangle \equiv|\psi(\lambda_{\theta},v)\rangle,
\end{equation}
from the
Gowdy states spanned in the basis of Bianchi I states $|\Psi(\lambda_{\theta},\lambda_{\sigma},
v)\rangle$ to the LRS-Gowdy states spanned in the basis of LRS-Bianchi I states
$|\psi(\lambda_{\theta},v)\rangle$. Here the sum in $\lambda_{\sigma}$ is carried out over the
considered superselection sector. A similar map was introduced in \cite{awe} to project from the
loop quantized Bianchi I model to its isotropic sector, the flat FRW model. Note that
in our case we cannot sum over $\lambda_\theta$ because the interchange $\theta \longleftrightarrow
\sigma$ is not a symmetry.

\section{Conclusions}
We have obtained a complete quantization of the Gowdy $T^{3}$ model with linearly polarized
gravitational waves and a minimally coupled inhomogeneous massless scalar field as matter content
using hybrid techniques in the framework of LQC. This hybrid quantization follows exactly in the
same way as that of the vacuum model because the matter and gravitational inhomogeneities can be
treated in the very same way. 

Owing to the polymeric quantization of the homogeneous gravitational sector we have been able to
eliminate the analog states to the classical cosmological singularities at the level of
superselection. Furthermore, the standard quantization of both matter and gravitational
inhomogeneities is recovered in physical states.  

The inclusion of the massless scalar field not only has allowed us to study matter
inhomogeneities in the framework of LQC by means of this hybrid quantization, but also leads to a
more physical interesting model. In fact, the homogeneous sector of the model admits now flat
Friedmann-Robertson-Walker (FRW) solutions.

This hybrid Gowdy $T^{3}$ model, or more specifically the simpler hybrid LRS-Gowdy $T^{3}$ model,
provides a perfect scenario for further study of the quantum back-reaction between anisotropies and
inhomogeneities and a loop quantized flat FRW background, as well as for the study of approximated methods in
LQC in order to extract physical predictions from the theory \cite{lrsg}.

\ack
This work was supported by the Spanish MICINN Projects No. FIS2008-06078-C03-03,
No. FIS2011-30145-C03-02, and the Consolider-Ingenio Program CPAN No. CSD2007-00042. D M-dB is
supported by CSIC and the European Social Fund under the Grant No. JAEPre-09-01796.

\end{document}